\documentclass[preprint,preprintnumbers,amsmath,amssymb,showkeys]{revtex4}
\usepackage{graphicx}

\newcommand{\abs}[1]{\left\vert#1\right\vert}

\newcommand{\amp}[2]{\left<#1 \vert #2 \right>}
\newcommand{\opamp}[3]{\left<#1 \left\vert #2 \right\vert #3 \right>}

\begin{document}
\title{Transmission spectra of two-dimensional quantum structures}
\author{Kia Manouchehri}
%\email{kia@physics.uwa.edu.au}
\author{J.B. Wang}
\email{wang@physics.uwa.edu.au} \affiliation{School of Physics,
The University of Western Australia}
\date{\today}
\begin{abstract}
To study the ballistic transport of charge carriers in
nano-structured quantum devices, a highly efficient numerical
technique is developed, which provides continuous transmission
spectra for arbitrarily complex potential geometries in two
dimensions. We apply the proposed method to single and double
barrier structures and compare the results with those obtained
using standard techniques for computing transmission coefficients.
Excellent numerical agreement as well as considerable
computational saving is demonstrated.
\end{abstract}


%\pacs{73.23.-b, 73.23.Ad, 73.23.Hk, 02.70.-c}

\maketitle

\section{Introduction}

A wide range of potential applications for nano-scale electronics
in the quantum regime, together with a rapid increase in computing
power, have generated much interest in the numerical analysis of
nano-structured devices as a viable means for studying their
properties \cite{27,26,25,4}. Examples of such applications based
on the transport properties of ballistic charge-carriers include
quantum wires \cite{10,11,12,13,14}, quantum resistors \cite{15},
resonant tunnelling diodes and band filters \cite{16,17}, quantum
transistors and stub tuners \cite{18,19,20}, quantum switches
\cite{21}, quantum sensors \cite{22,23,24}, as well as qubits and
quantum logic gates \cite{6,7}.

Much insight into the behavior of such devices can be gained by
studying the energy dependance of transmission through various
nano-structures for propagating charge carriers. The transmission
coefficient of a propagating wave function can be determined by
the application of time-independent (static) or time-dependent
(dynamic) techniques \cite{8}. The time-independent approach
involves solving the time-independent Schr\"{o}dinger equation,
commonly carried out using one of three methods: (1) Mode matching
method which is based on splitting the nano-structures into
different regions with known analytical solution. The total wave
function is then derived by matching the solutions at the boundary
between regions. This method has been widely used to study quantum
devices with simple geometries \cite{28,29,30,31}. In principle,
any arbitrary structure can be divided into many segments, each of
which is approximated as a rectangular potential well with a
finite width. For complex geometries however, this can result in a
very large matrix to be inverted and the computation can become
unstable. (2) Finite element method \cite{32} where using an
irregular numerical mesh, the potential and the wave function are
discretized by choosing a suitable set of basis functions to
approximate each grid point. However, due to the approximate
nature of the elemental solutions, large grids are still required
for accurately representing the wave function particularly when
dealing with excited or continuous energy states. This often
results in an excessive computational cost as well as the
introduction of high frequency noise, which can considerably
distort the phase of the wave function. (3) Green's function
method \cite{33,34} where the Schr\"{o}dinger's equation is
expressed and solved using the common Green's function techniques.
Although in principle capable of providing solutions to
arbitrarily complex problems independently of the potential
boundary structures, the Green's function method is most efficient
when used in conjunction with simplifying assumptions such as the
decoupling of the $x$ and $y$ variables in the two-dimensional
Schr\"{o}dinger equation or by not requiring information about a
global wave function. In particular, the later case leads to an
efficient algorithm for calculating the transmission coefficients
known as the recursive Green's function method \cite{35}, where it
is necessary to compute the Green's function only for some
sections of the structure. The calculation of the wave function
for a general two-dimensional case however relies on computing the
generalized Green's function for the entire region, which is
computationally expensive.

As well as their limited utility in constructing global wave
functions for complex nano-structure geometries efficiently, the
time-independent methods are also unable to provide information on
the transient behavior of the system under study. This can be
addressed by solving the time-dependent Schr\"{o}dinger equation
where transmission coefficients are obtained by numerically
propagating an initial wave function through a given potential
geometry and then summing over the transmitted parts. Methods for
solving the time-dependent Schr\"{o}dinger equation include: (1)
The finite difference method \cite{36,37}, which uses an
\emph{n}th order finite difference approximation to expand the
time-evolution operator. This method however suffers from a low
convergence rate as well as excessive truncation errors over
modestly long propagation times. (2) Split operator method
\cite{38}, where the time-evolution operator is approximated as
the product of three diagonal operators which can be readily
solved. This scheme however neglects the commutator between these
operators and thus introduces error in both energy and phase of
the wave function. (3) The time-dependent Green's functions method
\cite{39}, which employs the standard Green's function techniques.
Due to its reliance on summations over a large number of
energy-eigenstates however, this method too becomes
computationally prohibitive when dealing with complex potentials.
(4) Chebyshev-Fourier propagation scheme, which is based on
expressing the evolution operator using the Chebyshev
approximation \cite{9}. This method proves to be numerically
robust as well as efficient for propagations over arbitrary
potentials.

Nonetheless, when approached naively, calculating the transmission
coefficients for a wide range of energies using the
Chebyshev-Fourier scheme could still prove computationally
expensive, amounting to repeated propagations of an initial wave
function with various mean energies in the given range. A novel
method for computing the transmission coefficients was developed
by Yin and Wang \cite{2}, which arrives at the entire transmission
spectrum for a given range of energies after only one propagation
of an initial wave function with a broad energy distribution.
Although this method is shown to be highly efficient while
maintaining the capacity to deal with arbitrary potentials in one
dimension, its use in two dimensions is limited to geometries with
infinite boundary conditions in the second dimension and decoupled
$x$ and $y$ variables \cite{2,3}.

In this paper we extend the work of Yin and Wang \cite{2}, making
it possible to treat arbitrarily complex potentials in two
dimensions, without imposing any limitations on the dimensional
parameters of the system, such as decoupling in the $x$ and $y$
directions. We apply this scheme to single and double hyperbolic
barrier structures and demonstrate excellent numerical accuracy
while maintaining a high degree of computational efficiency,
compared to standard techniques for obtaining individual
transmission coefficients.

\section{Theory}

We begin by considering the two-dimensional Schr\"{o}dinger
equation
\begin{equation}
    i \hbar \frac{\partial}{\partial t} \psi(x,y,t) = \hat{H} \psi(x,y,t),
\end{equation}
where the Hamiltonian $\hat{H}$ is given by
\begin{equation}
\hat{H} = - \frac{\hbar^{2}}{2m^{\ast}} \nabla^{2} + V(x,y,t)
\end{equation}
with $m^{\ast}$ being the effective mass of the propagating charge
carrier within the semiconducting lattice. If the potential
function is time-independent i.e $V(x,y,t) \equiv V(x,y)$, then a
formal solution to this equation is
\begin{equation}\label{eqn.schrodinger}
    \psi(x,y,t) = exp(-\frac{i} {\hbar} \hat{H} t) \psi(x,y,0)
\end{equation}
where $t$ is the propagation time, and $exp(-\frac{i} {\hbar}
\hat{H} t)$ is the time evolution propagator, commonly denoted by
$\hat{U}$. The Chebyshev-Fourier scheme as detailed in \cite{40,9}
approximates $\hat{U}$ by a Chebyshev polynomial expansion
\begin{equation}
    \psi(x,y,t)=\exp[-i(\mathcal{E}_{max}+\mathcal{E}_{min})t]
    \sum_{n=0}^\mathcal{N}a_n(\alpha)\phi_n(-\mathcal{\widetilde{H}})\psi(x,y,0)
\end{equation}
where $\mathcal{E}_{min}$ and $\mathcal{E}_{max}$ are the upper
and lower bounds on the energies sampled by the wave packet,
$a_n(\alpha)= 2J_n(\alpha)$ except for $a_0(\alpha)=J_0(\alpha)$,
$J_n(\alpha)$ are the Bessel functions of the first kind, $\phi_n$
are the Chebyshev polynomials, and $\mathcal{N}$ is the number of
terms in the Chebyshev expansion. To ensure convergence, the
Hamiltonian needs to be normalized as
\begin{equation}\label{eqn.Chebyshev}
    \mathcal{\widetilde{H}}=\frac{1}{\mathcal{E}_{max}-\mathcal{E}_{min}}
    [2\hat{H}-\mathcal{E}_{max}-\mathcal{E}_{min}].
\end{equation}
The action of the Laplacian operator $\nabla$ on the wave
functions is carried out using a Fourier transformation technique
\cite{40}. It is important to note that the rapid convergence and
the high numerical accuracy of this propagation scheme are well
established in the literature \cite{1,9,5,4,40}, where it has been
used in the study of various quantum nano-structures.

\begin{figure}[h]
    \centering
    \includegraphics[width=7cm]{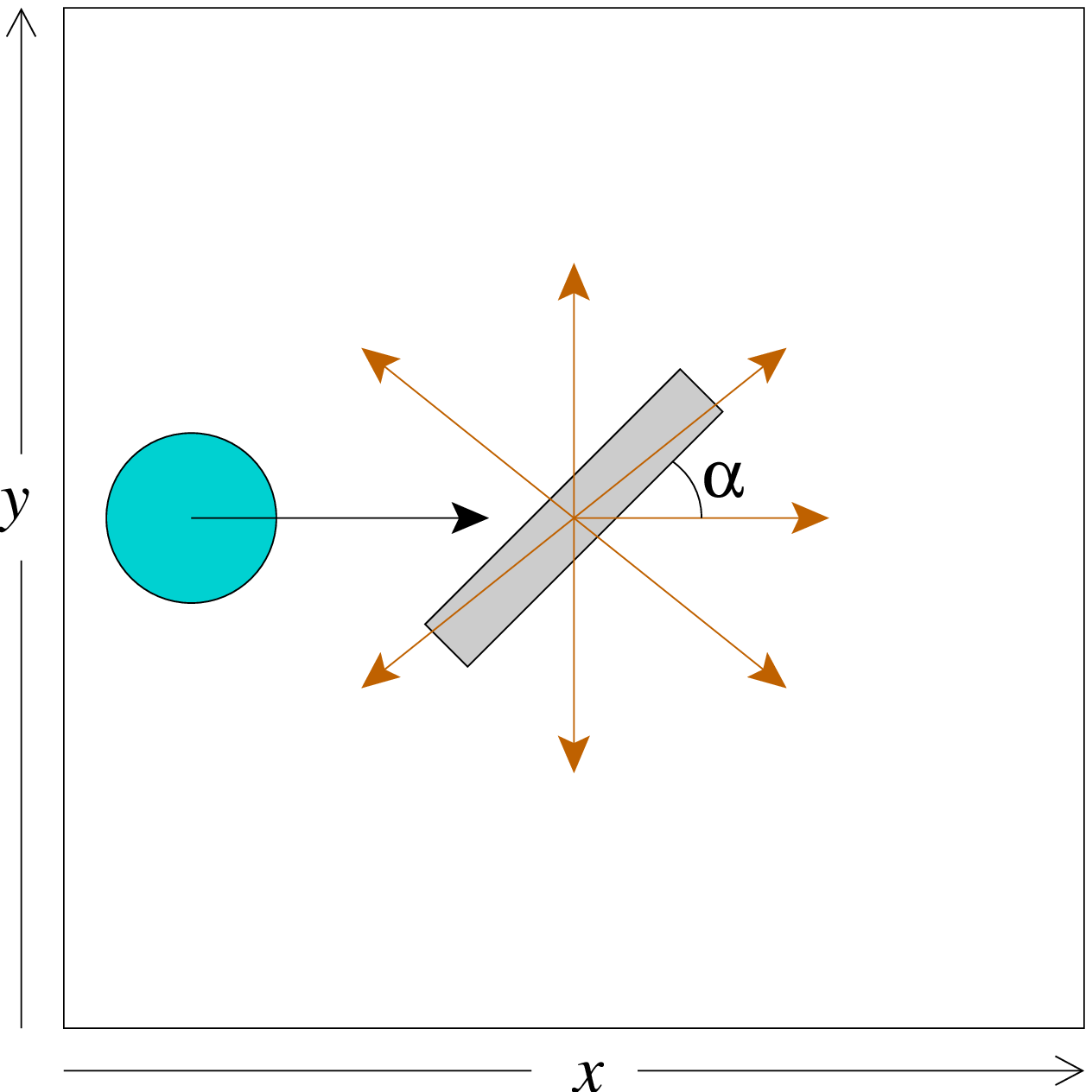}
    \caption{Position space: the incident particle approaches the
    scattering potential barrier from the left in the $x$ direction.
    The angle $\alpha$ describes the orientation of the
    barrier with respect to the incident particle.}
    \label{fig.space-x}
\end{figure}

\begin{figure}[h]
    \centering
    \includegraphics[width=9cm,bbllx=120,bblly=160,bburx=470,bbury=520,clip]{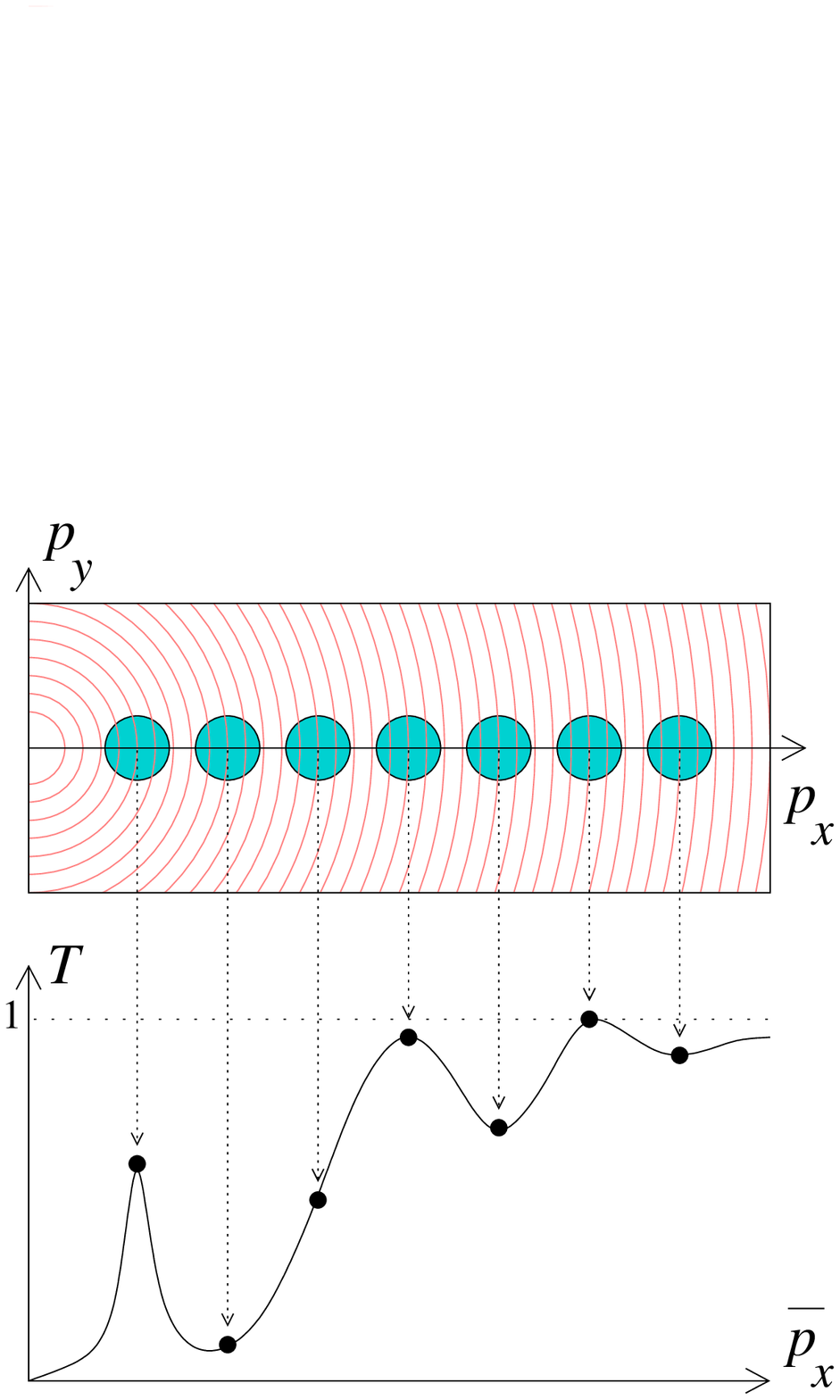}
    \caption{In the momentum space incident charge carriers are modelled using initial
    wave packets with a narrow momentum spread, positioned at
    $\overline{p}_{x}=\sqrt{2m^{\ast}\overline{\mathcal{E}}}$ and
    $\overline{p}_{y}=0$, where $\overline{\mathcal{E}}$
    is the mean energy of the wave packet.
    A complete transmission curve can be constructed by computing the
    transmission coefficient of every
    wave packet with $\overline{\mathcal{E}}$ in the given range.}
    \label{fig.pwfn-a}
\end{figure}

In the simulations we start with a localized wave packet as the
initial wave function $\psi_{i}=\psi(x,y,t=0)$ and then use the
Chebyshev-Fourier scheme as a means to propagate this wave packet
in time $t$. The final wave function $\psi_{f}=\psi(x,y,t=\tau)$
is partly transmitted and partly reflected due to the potential
barrier $V(x,y)$ (Fig. \ref{fig.space-x}). To evaluate the
transmission coefficient $T(\overline{\mathcal{E}})$ for any given
potential barrier, one can setup $\psi_{i}$ with an incident mean
energy $\overline{\mathcal{E}}$ and a very small energy
uncertainty, propagate $\psi_{i}$ in time until it is completely
scattered by the barrier, and then collect the transmitted parts
of $\psi_{f}$. In this way, obtaining a transmission spectrum
entails multiple propagations for different incident energies as
depicted in Fig. \ref{fig.pwfn-a}. Given the computational cost
associated with any time-dependant propagation scheme however, the
use of this \emph{direct method} for obtaining a continues
transmission spectrum is clearly inefficient.

In Yiu and Wang's approach \cite{2}, a wave packet with a small
spatial uncertainty (corresponding to a large momentum spread) is
used to obtain the entire transmission carve after only one
propagation. In other words, one can compute the transmission
coefficients by simply transforming the transmitted wave packet to
momentum space and dividing by the original momentum space wave
function. Noting that in one dimension, the transmitted and
reflected wave packets can be expressed as momentum space
components of $\psi_{f}$ with positive momentum $(p>0)$ and
negative momentum $(p<0)$ respectively, one can then write
\begin{equation}\label{eqn.1}
    T(p) = \frac{\abs{\psi_{f}(p>0)}^{2}}{\abs{\psi_{i}(p)}^{2}} \text{~~~and~~~}
    R(p) = \frac{\abs{\psi_{f}(p<0)}^{2}}{\abs{\psi_{i}(p)}^{2}} \text{~,}
\end{equation}
where $T(p)$ and $R(p)$ are the transmission and reflection
coefficients at momentum $p$. This scheme however, is strictly
limited to scattering in one dimension. In order to show this we
present a more generalized approach to the problem, which enables
us to extend the current scheme to two dimensions.

First we note that in principal $\psi(x,y,t)$ may be expressed as
a linear superposition of energy eigenfunctions $\phi_{E}(x,y)$.
We can therefore rewrite Eq. (\ref{eqn.schrodinger}) as

\begin{eqnarray}\label{eqn.schrodinger.energy}
    \psi(x,y,t)&=&e^{-i\hat{H}t/\hbar}\sum_{E}\!\!\!\!\!\!\!\!\int c_{i}(E)\phi_{E}(x,y)\\
               &=&\sum_{E}\!\!\!\!\!\!\!\!\int e^{-iEt/\hbar}c_{i}(E)\phi_{E}(x,y) \\
               &=&\sum_{E}\!\!\!\!\!\!\!\!\int c_{f}(E)\phi_{E}(x,y) \text{~,}
\end{eqnarray}

where $c_{i}(E)$ and $c_{f}(E)$ denote the initial and final
complex coefficients associated with each energy eigenfunction. We
note that $c_{i}(E)$ and $c_{f}(E)$ only differ by a phase factor.
Expressing $c_{f}(E)$ as $\amp{E}{\psi_{f}}$ we can expand
$c_{f}(E)$ using the position basis set
\begin{eqnarray}
    c_{f}(E)&=&\iint\amp{E}{x,y}\amp{x,y}{\psi_{i}}\, dx\, dy\\
    \abs{c_{f}(E)}^{2} &=&
    \iint\amp{E}{x,y}\amp{x,y}{\psi_{f}}\amp{\psi_{f}}{E}\, dx\, dy
\end{eqnarray}
In the position space however the final wave function assumes two
components: the transmitted part lying in an area $T$, and the
reflected part lying in an area $R$. Noting that these two areas
do not overlap we can write
\begin{eqnarray}
    \abs{c_{f}(E)}^{2} &=&
    \iint_{T}\amp{E}{x,y}\amp{x,y}{\psi_{f}}\amp{\psi_{f}}{E} \, dx\, dy\\
    &+&\iint_{R}\amp{E}{x,y}\amp{x,y}{\psi_{f}}\amp{\psi_{f}}{E}\, dx\, dy\\
    &=&\abs{c_{T}(E)}^{2} + \abs{c_{R}(E)}^{2}
\end{eqnarray}
where $c_{T}(E)$ and $c_{R}(E)$ are the transmission and
reflection coefficients of each energy eigenfunction. We therefore
conclude that
\begin{equation}\label{eqn.k.conservation}
                \abs{c_{i}(E)}^{2} =
                \abs{c_{f}(E)}^{2} =
                \abs{c_{T}(E)}^{2} + \abs{c_{R}(E)}^{2}.
\end{equation}
In other words the amplitude of each energy eigenfunction will
independently conserve throughout the propagation and must
therefore have its own transmitted and reflected parts.

In order to evaluate $\abs{c_{i}(E)}^2$, $\abs{c_{T}(E)}^2$ and
$\abs{c_{R}(E)}^2$, we consider the case where the initial and the
scattered wave functions are both sufficiently distant from the
potential barrier, such that the interaction between the wave
function and the potential barrier is negligible, i.e.
$\opamp{\psi}{\hat{V}}{\psi} \rightarrow 0$. As a result we can
ignore the contribution of the potential term $V(x,y)$ in the
Hamiltonian since $\hat{H}\rightarrow\hat{p}^{2}/2m^{\ast}$,
meaning that energy is directly proportional to the square of the
momentum operator.

\begin{figure}[h]
    \centering
    \includegraphics[width=7cm]{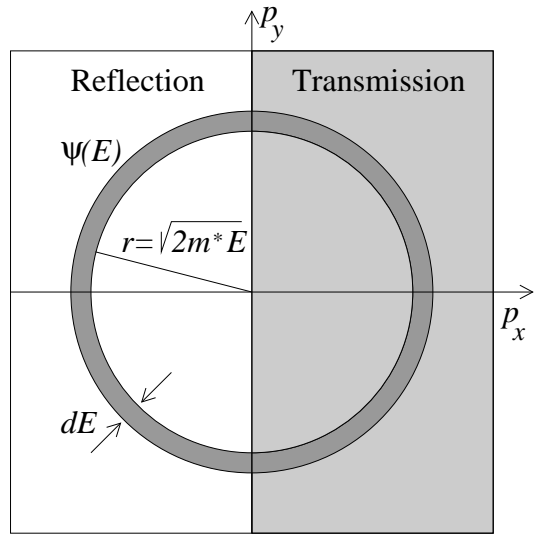}
    \caption{Momentum space: transmission and reflection are
    defined as wave components with $p_{x} > 0$ and $p_{x} < 0$ respectively.
    Components $\psi(E)$ correspond to rings of radius $r=\sqrt{2m^{\ast}E}$.}
    \label{fig.space-p}
\end{figure}

Now representing the initial wave functions in the momentum space,
denoted by $\psi_{i}(p_{x},p_{y})=\frac{1}{2\pi\hbar}\int\int
\exp(\frac{-i}{\hbar} (p_{x} x + p_{y} y)) \psi_{i}(x,y) dx dy$,
and noting that the total momentum $p^{2}=p_{x}^{2}+p_{y}^{2}$, we
have
\begin{equation}
    \abs{c_{i}(E)}^{2} =
    \sum_{p_{x}}\!\!\!\!\!\!\!\!\int\sum_{p_{y}}\!\!\!\!\!\!\!\!\int\abs{\psi_{i}(p_{x},p_{y})}^{2}\text{,~~for~}
    p_{x}^{2} + p_{y}^{2} = 2m^{\ast}E.
\end{equation}
In this way computing $\abs{c_{i}(E)}^{2}$ amounts to simply
summing over those momentum components which lie on a circle with
radius $\sqrt{2m^{\ast}E}$ as depicted in Fig. \ref{fig.space-p}.
Furthermore, by defining the motion of the initial wave packet to
be in the positive $x$ direction (see Fig. \ref{fig.space-x}),
transmission and reflection can also be conveniently characterized
using the momentum space representation of the final wave function
$\psi_{f}(p_{x},p_{y})$. Referring to the momentum circle
described above, $\abs{c_{T}(E)}^2$ and $\abs{c_{R}(E)}^2$
correspond to semicircles with $p_{x}>0$ and $p_{x}<0$
respectively. i.e.
\begin{equation}
    \abs{c_{T}(E)}^{2} =
    \sum_{p_{x}>0}\!\!\!\!\!\!\!\!\int\sum_{p_{y}}\!\!\!\!\!\!\!\!\int \abs{\psi_{f}(p_{x},p_{y})}^{2}\text{,~~for~}
    p_{x}^{2} + p_{y}^{2} = 2m^{\ast}E,
\end{equation}
\begin{equation}
    \abs{c_{R}(E)}^{2} =
    \sum_{p_{x}<0}\!\!\!\!\!\!\!\!\int\sum_{p_{y}}\!\!\!\!\!\!\!\!\int \abs{\psi_{f}(p_{x},p_{y})}^{2}\text{,~~for~}
    p_{x}^{2} + p_{y}^{2} = 2m^{\ast}E.
\end{equation}

\begin{figure}[h]
    \centering
    \includegraphics[width=9cm,bbllx=130,bblly=365,bburx=470,bbury=520,clip]{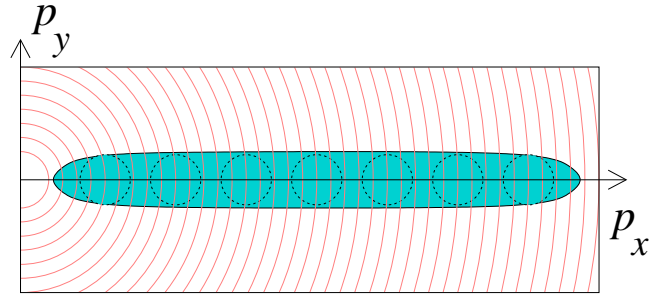}
    \caption{A broad-energy wave packet has a large momentum
    spread in the $x$ direction, corresponding to the energy range
    of the transmission spectrum. In the $y$ direction however, it
    has the same spread as the single charge carrier wave packets.}
    \label{fig.pwfn-b}
\end{figure}

In this way, the transmission coefficient $T(E)$ can be written as
\begin{equation}\label{eqn.transmission}
    T(E)=\frac{\abs{c_{T}(E)}^{2}}{\abs{c_{i}(E)}^{2}}\text{~.}
\end{equation}
Therefore, in principle, all the transmission coefficients can be
calculated after propagating only a single initial wave packet
$\psi_{i}$ for any choice of $\overline{\mathcal{E}}$. In this
work we have used an initial wave packet which has a small
momentum spread in the $y$ direction, but a broad momentum
distribution in the $x$ direction, as depicted in Fig.
\ref{fig.pwfn-b}. The scattering of this customized wave packet,
which we refer to as the \emph{broad-energy wave packet}, provides
us with an efficient means for computing the continuous
transmission spectrum over a wide range of incident energies for
an arbitrary potential barrier in two dimensions. We term this
scheme the \emph{momentum space method}.

\section{Results}

\begin{figure}[h]
    \centering
    \includegraphics[clip,width=9cm,bbllx=0,bblly=0,bburx=590,bbury=450]{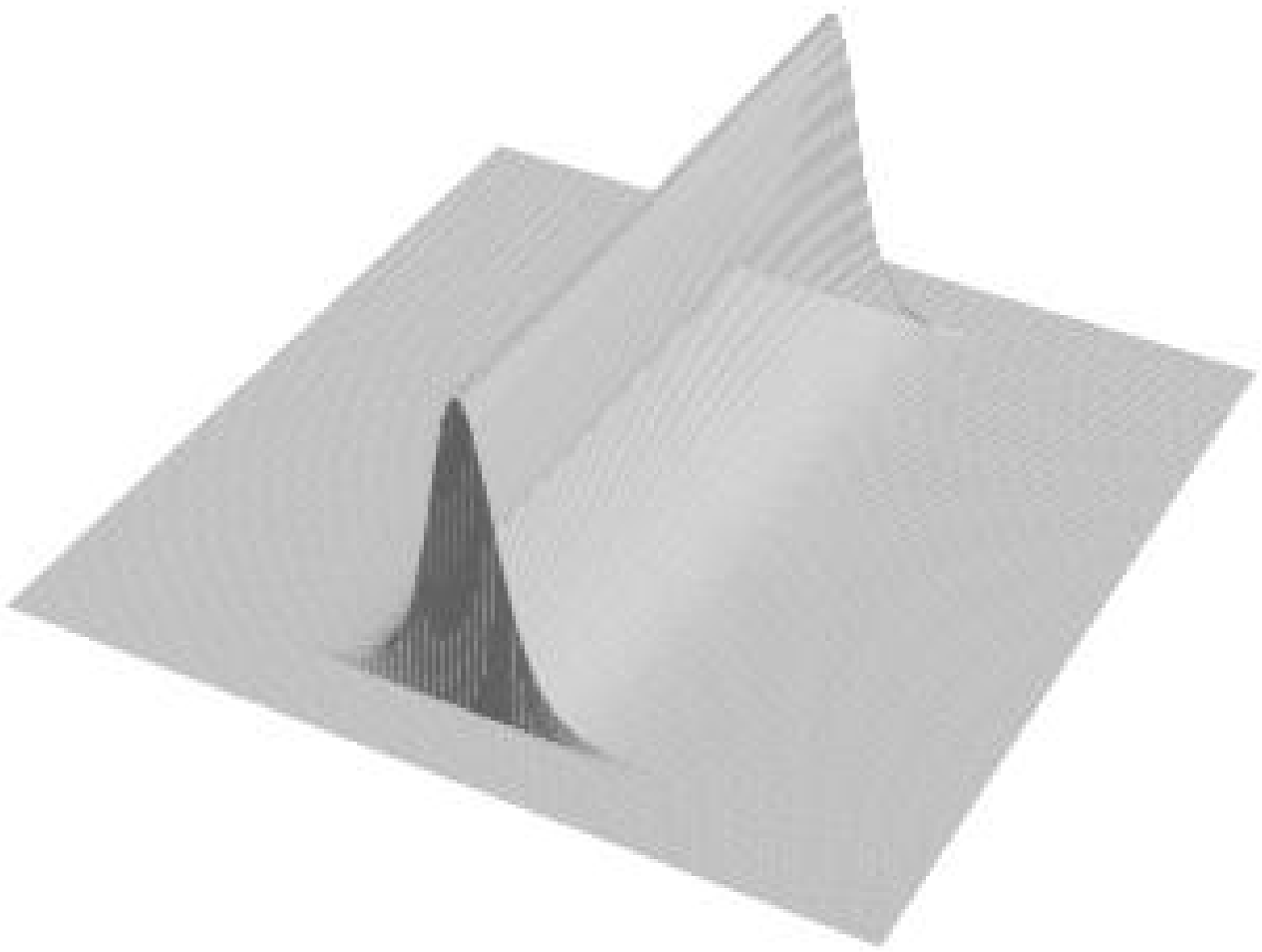}\\
    \includegraphics[clip,width=9cm,bbllx=0,bblly=0,bburx=590,bbury=450]{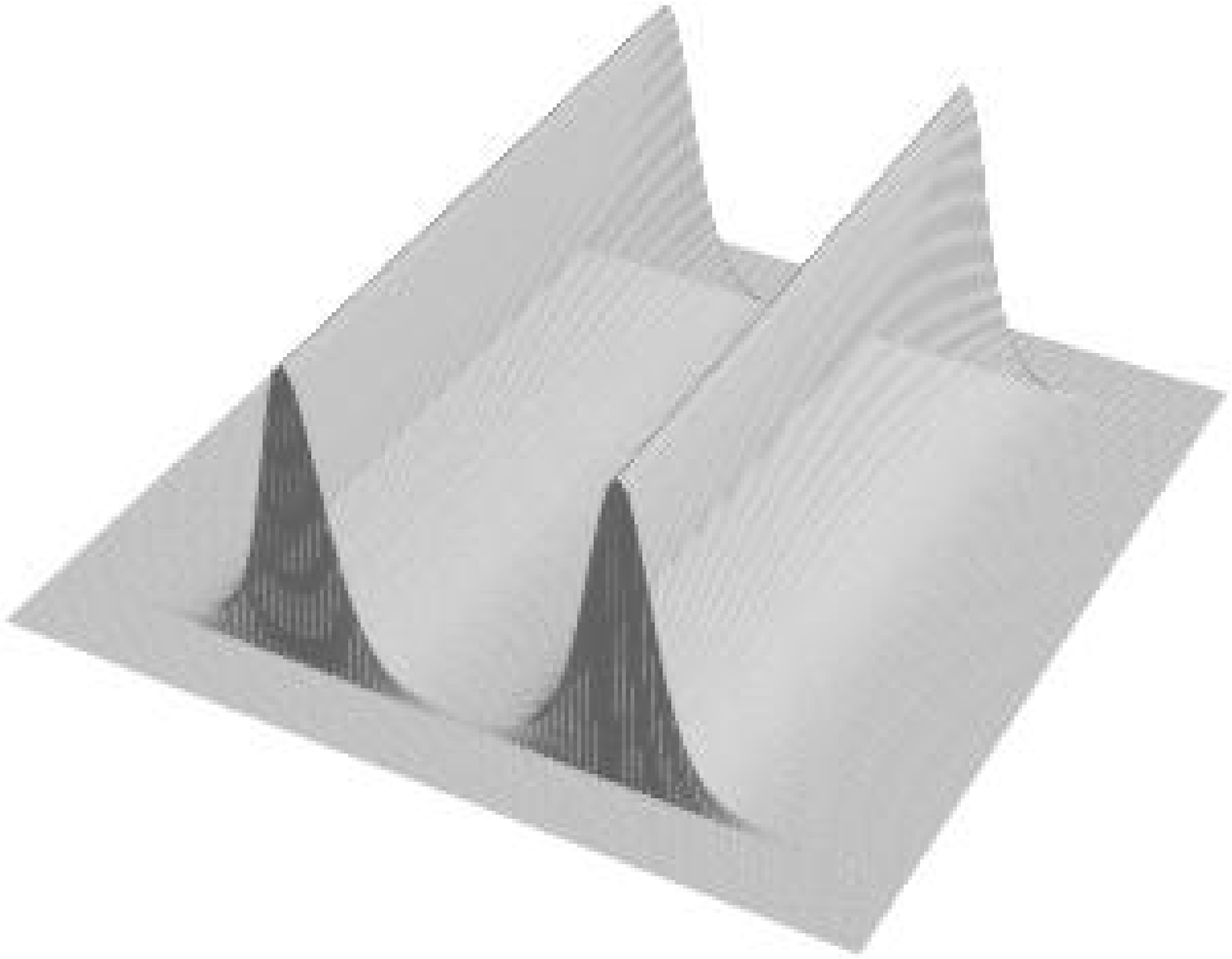}
    \caption{Single and double barrier potentials.}
    \label{fig.pot}
\end{figure}

In this section we demonstrate the use of the momentum space
technique to obtain a series of continuous transmission spectra
for the scattering of ballistic electrons by the single and double
barrier potentials (Fig.~\ref{fig.pot}), positioned at various
angles $\alpha$ with respect to the incident electron (Fig.
\ref{fig.space-x}). We then compare the results with transmission
coefficients obtained using the direct method (using multiple wave
packets with small momentum uncertainty).

The barriers are constructed using hyperbolic functions in the $x$
direction: $V(x) = V_{0}/\cosh^{2}(\frac{x}{a})$ for the single
barrier and $V(x) = V_{0}/\cosh^{2}(\frac{x-d}{a}) +
V_{0}/\cosh^{2}(\frac{x+d}{a})$ for the double barrier where the
barrier height $V_{0} = 13.6$ eV, width parameter $a = 0.21$ nm
(giving a half width of $0.19$ nm), and the barrier separation $2d
= 1.58$ nm. Simulations are carried out using three orientation
angles $\alpha=90^{\circ}, 62^{\circ},$ and $52^{\circ}$.

\begin{figure}[h]
    \centering
    \includegraphics[clip,width=6cm,bbllx=0,bblly=0,bburx=590,bbury=460]{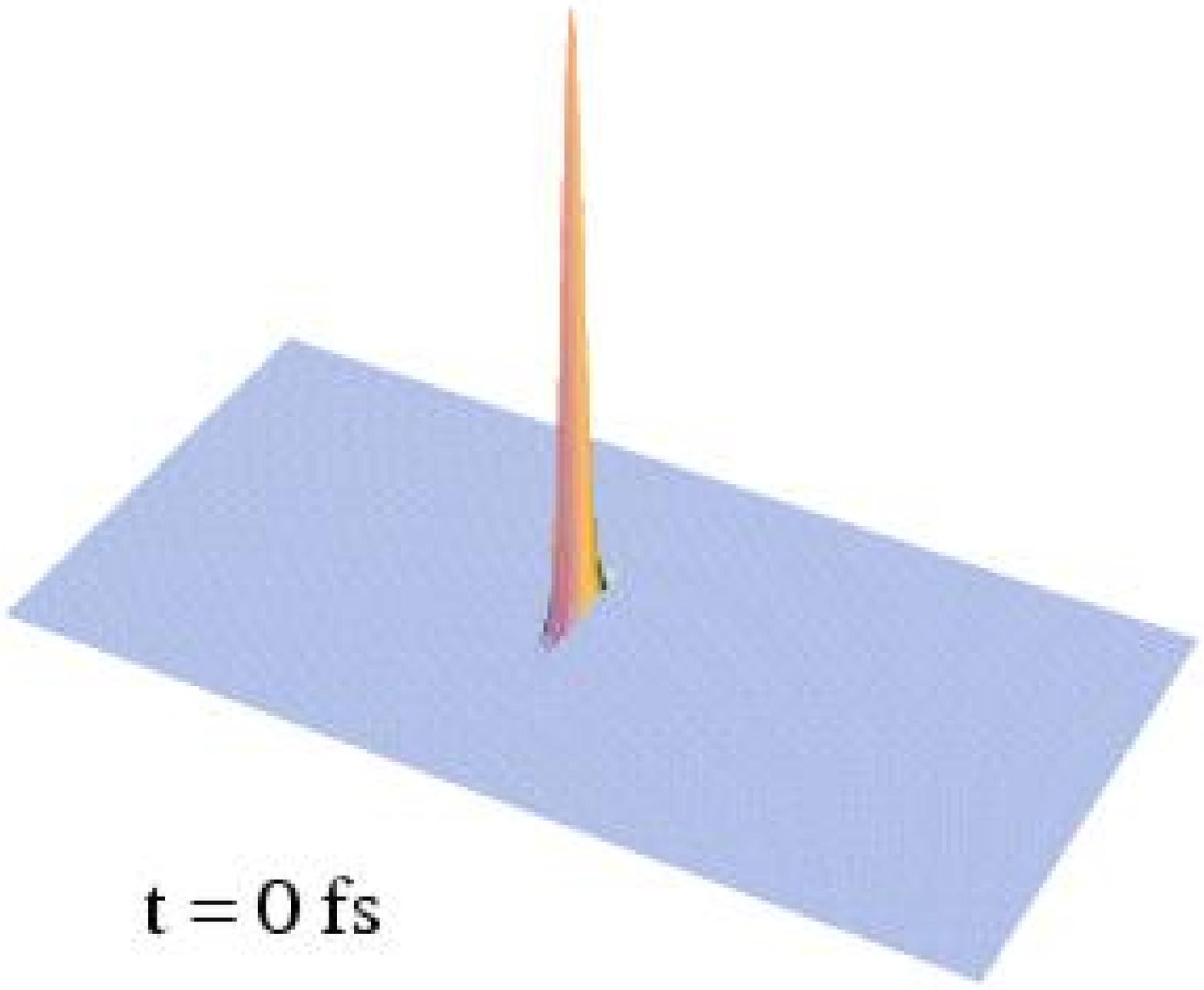}\\
    \includegraphics[clip,width=6cm,bbllx=0,bblly=0,bburx=590,bbury=460]{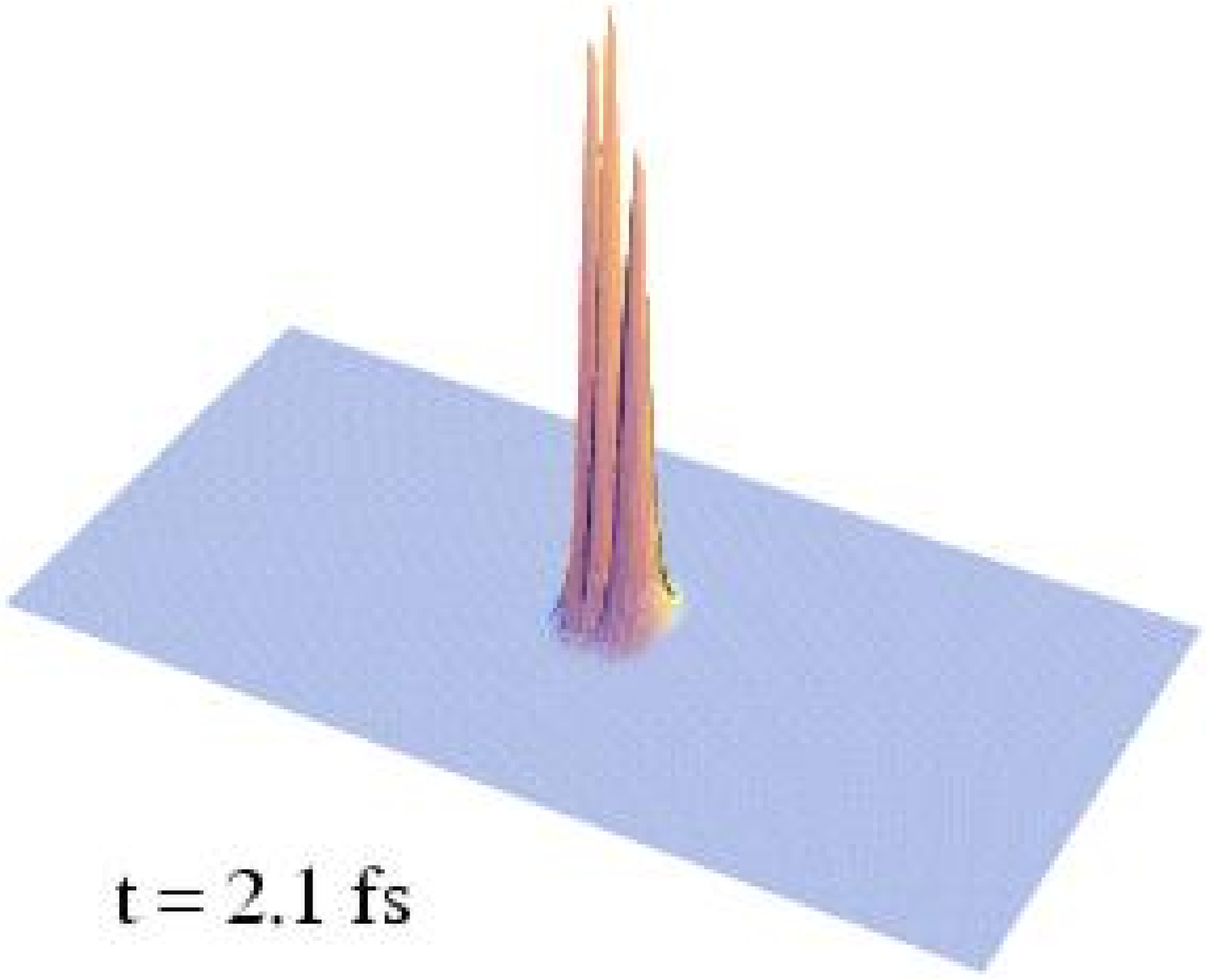}
    \includegraphics[clip,width=6cm,bbllx=0,bblly=0,bburx=590,bbury=460]{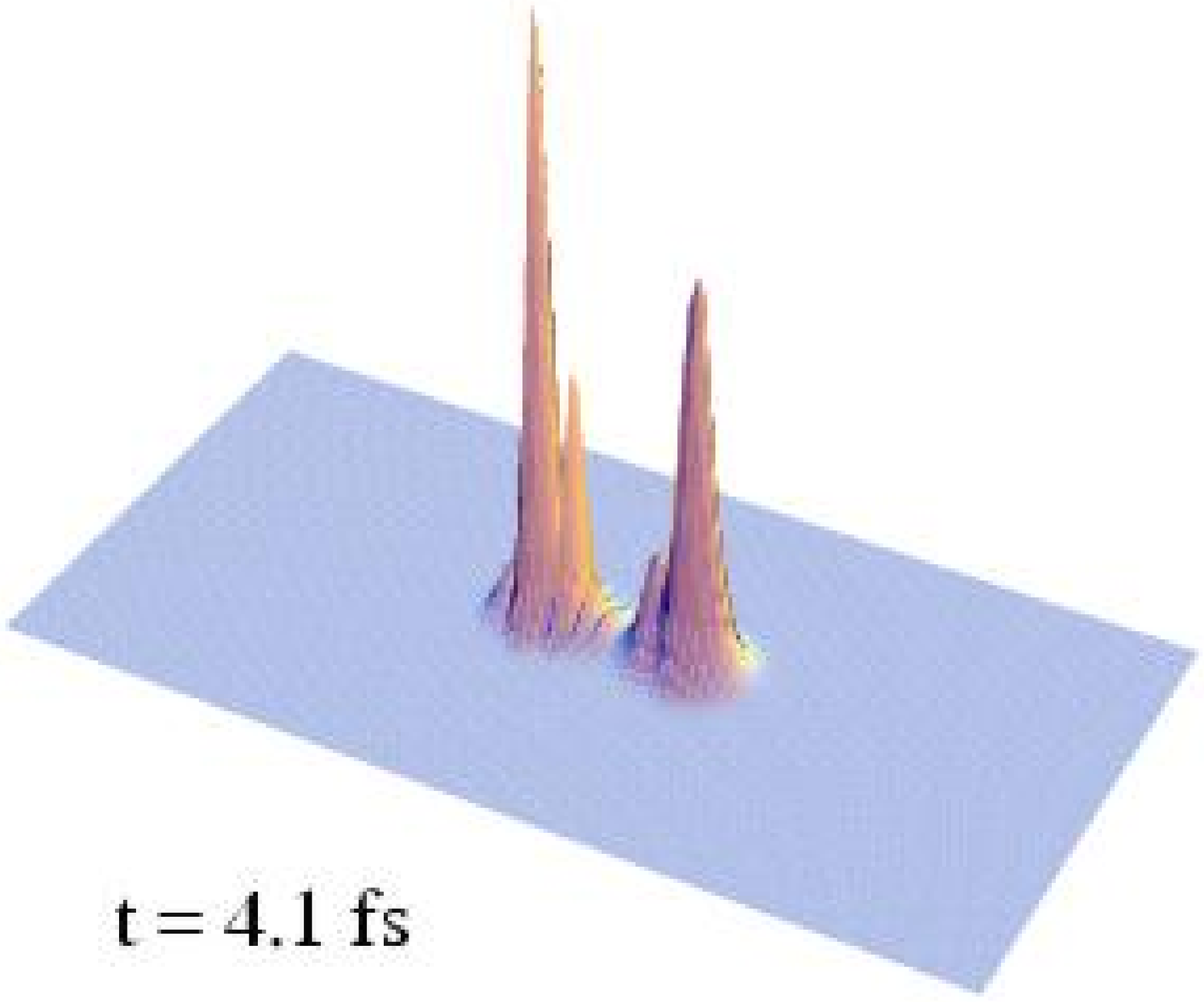}
    \includegraphics[clip,width=6cm,bbllx=0,bblly=0,bburx=590,bbury=460]{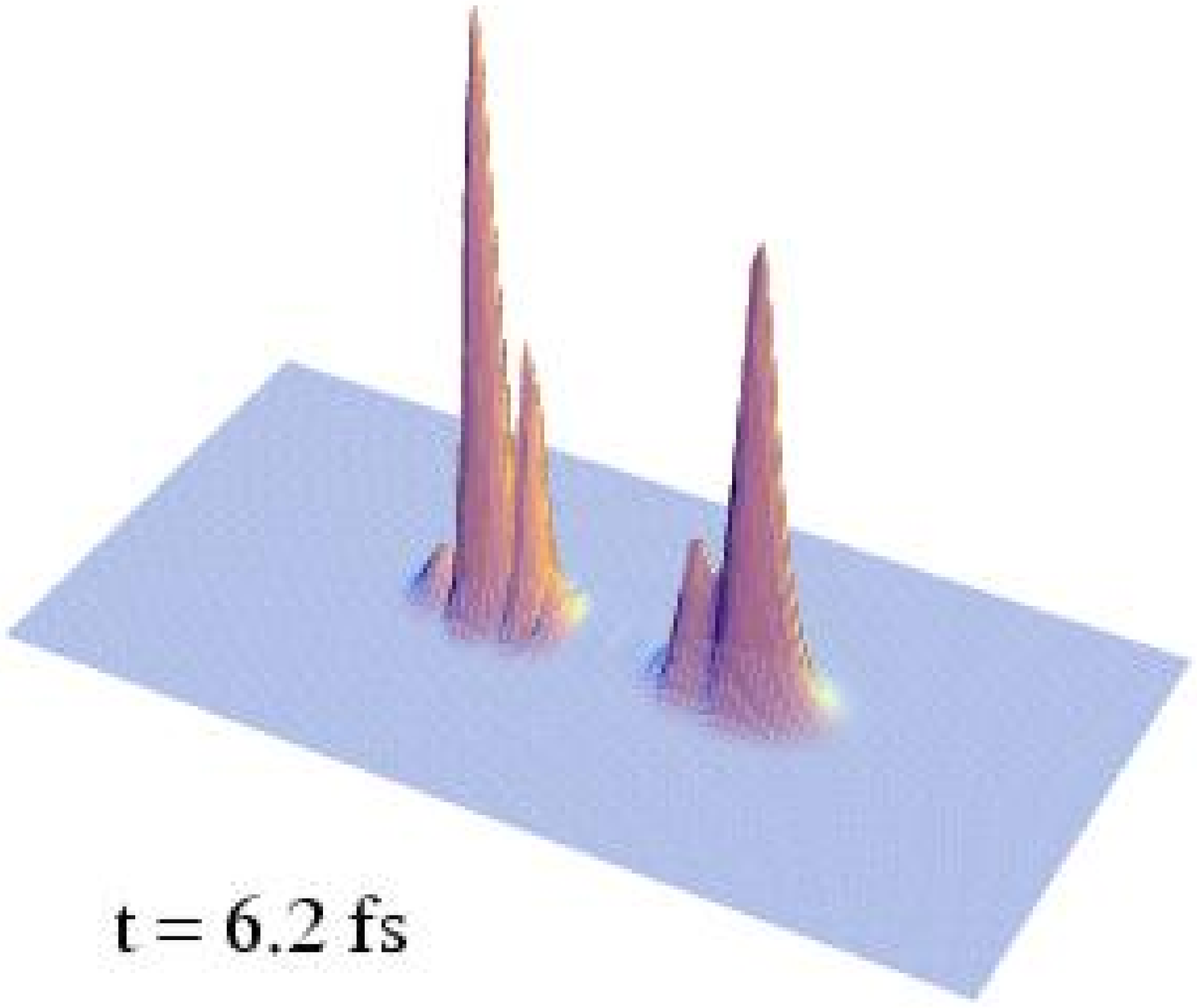}
    \includegraphics[clip,width=6cm,bbllx=0,bblly=0,bburx=590,bbury=460]{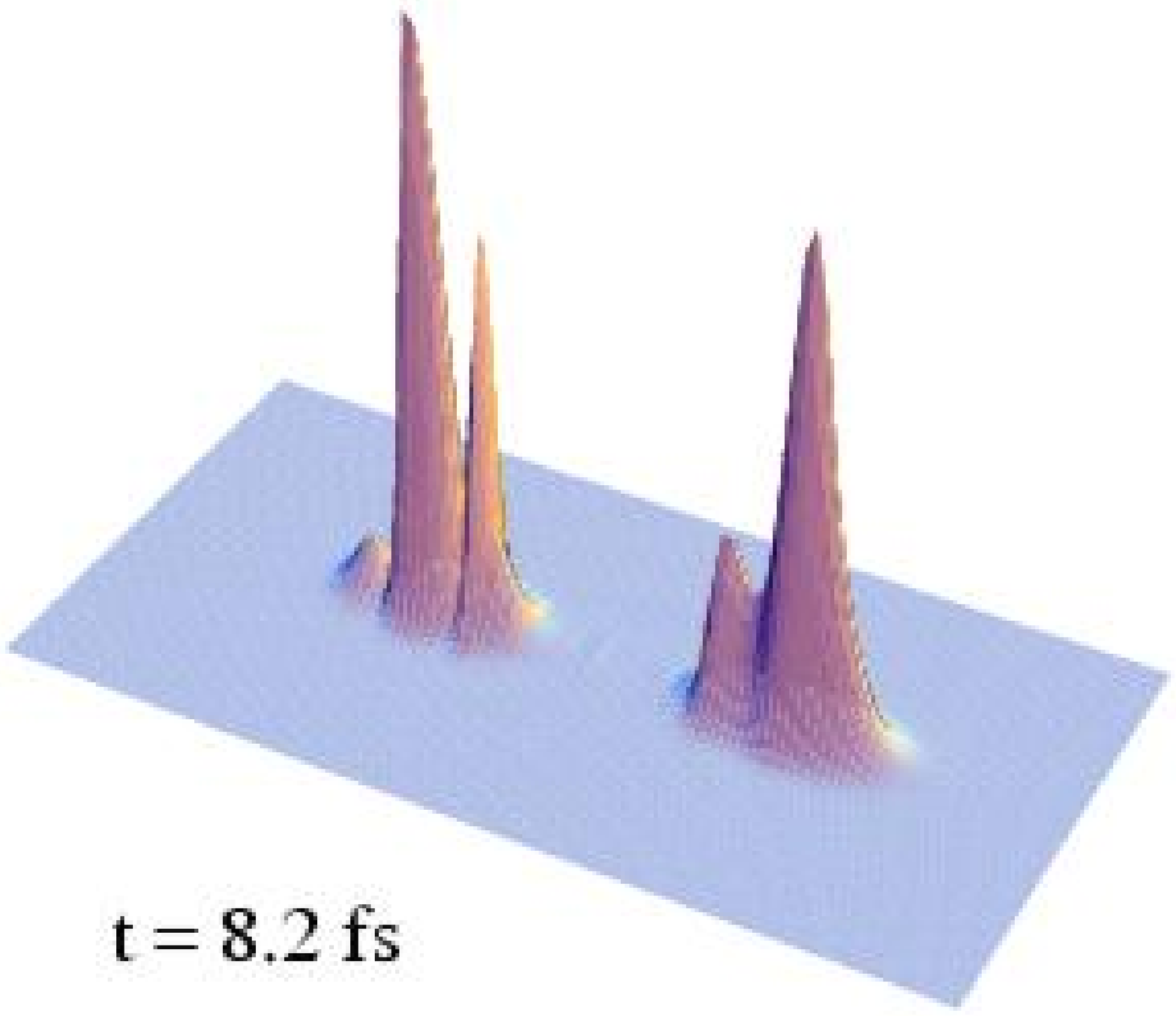}
    \includegraphics[clip,width=6cm,bbllx=0,bblly=0,bburx=590,bbury=460]{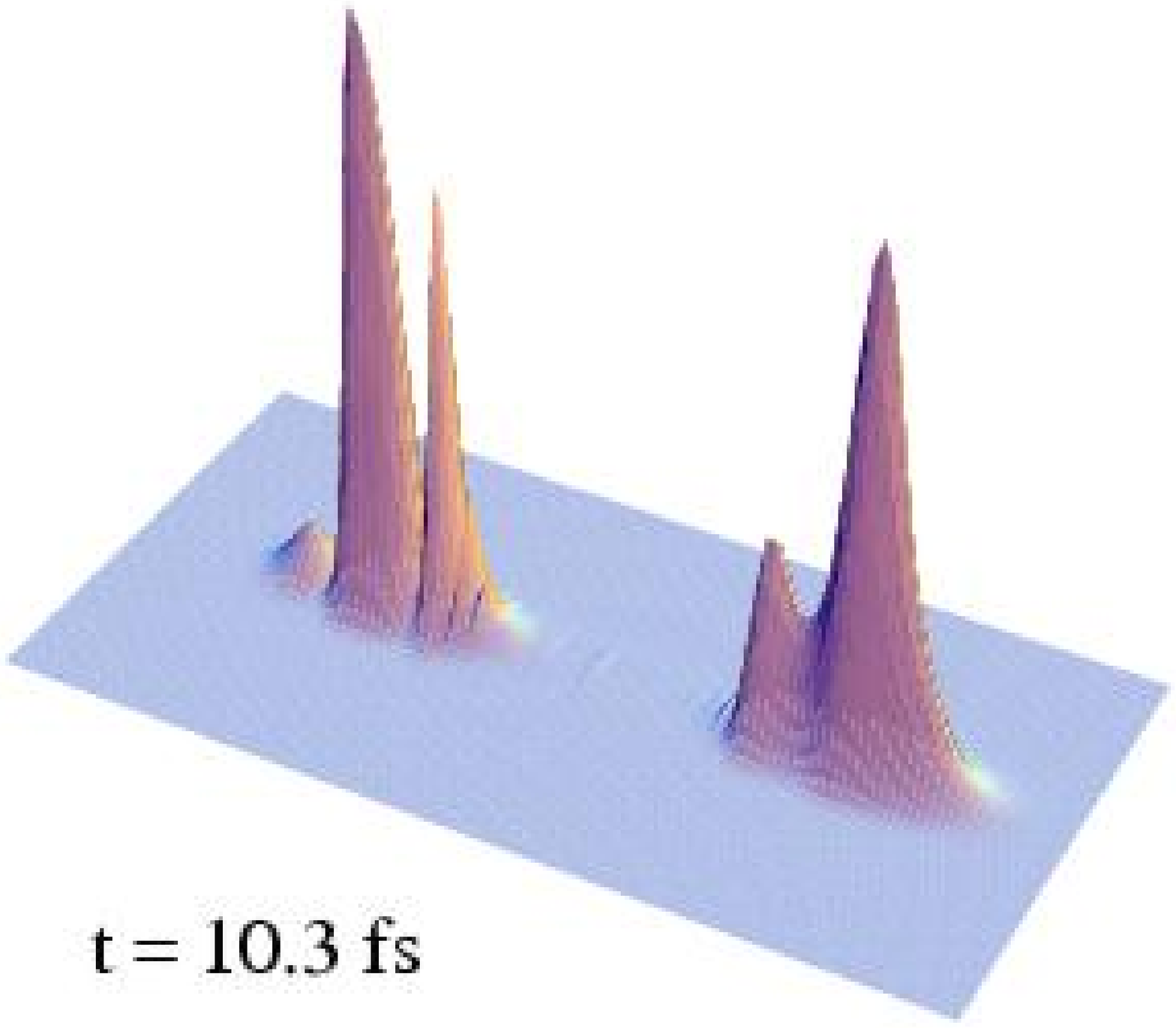}
    \caption{Snapshots of the broad-energy wave packet scattering from the double barrier potential
    with $\alpha=90^{\circ}$.}
    \label{fig.prop-00}
\end{figure}

\begin{figure}[h]
    \centering
    \includegraphics[clip,width=6cm,bbllx=0,bblly=0,bburx=590,bbury=480]{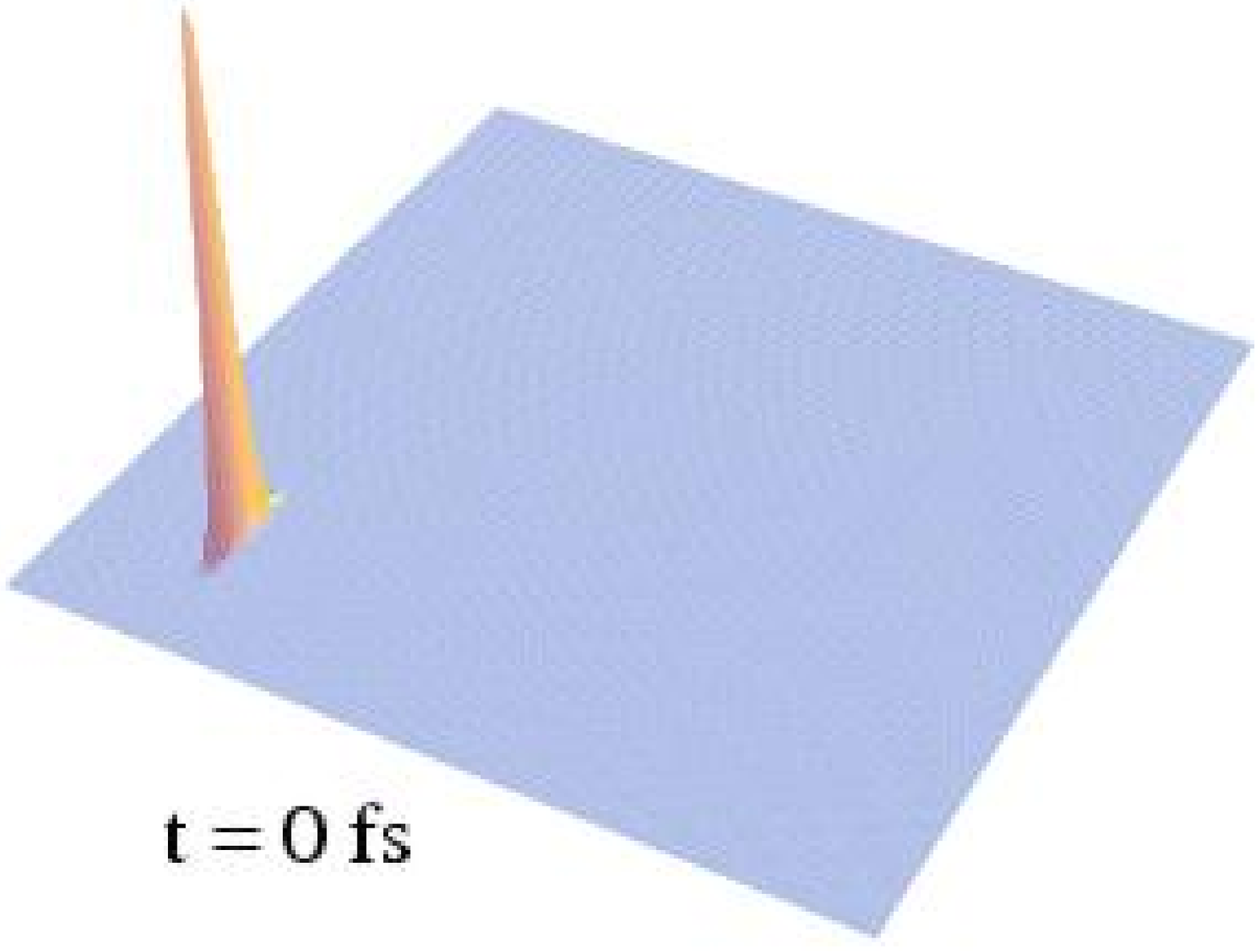}\\
    \includegraphics[clip,width=6cm,bbllx=0,bblly=0,bburx=590,bbury=480]{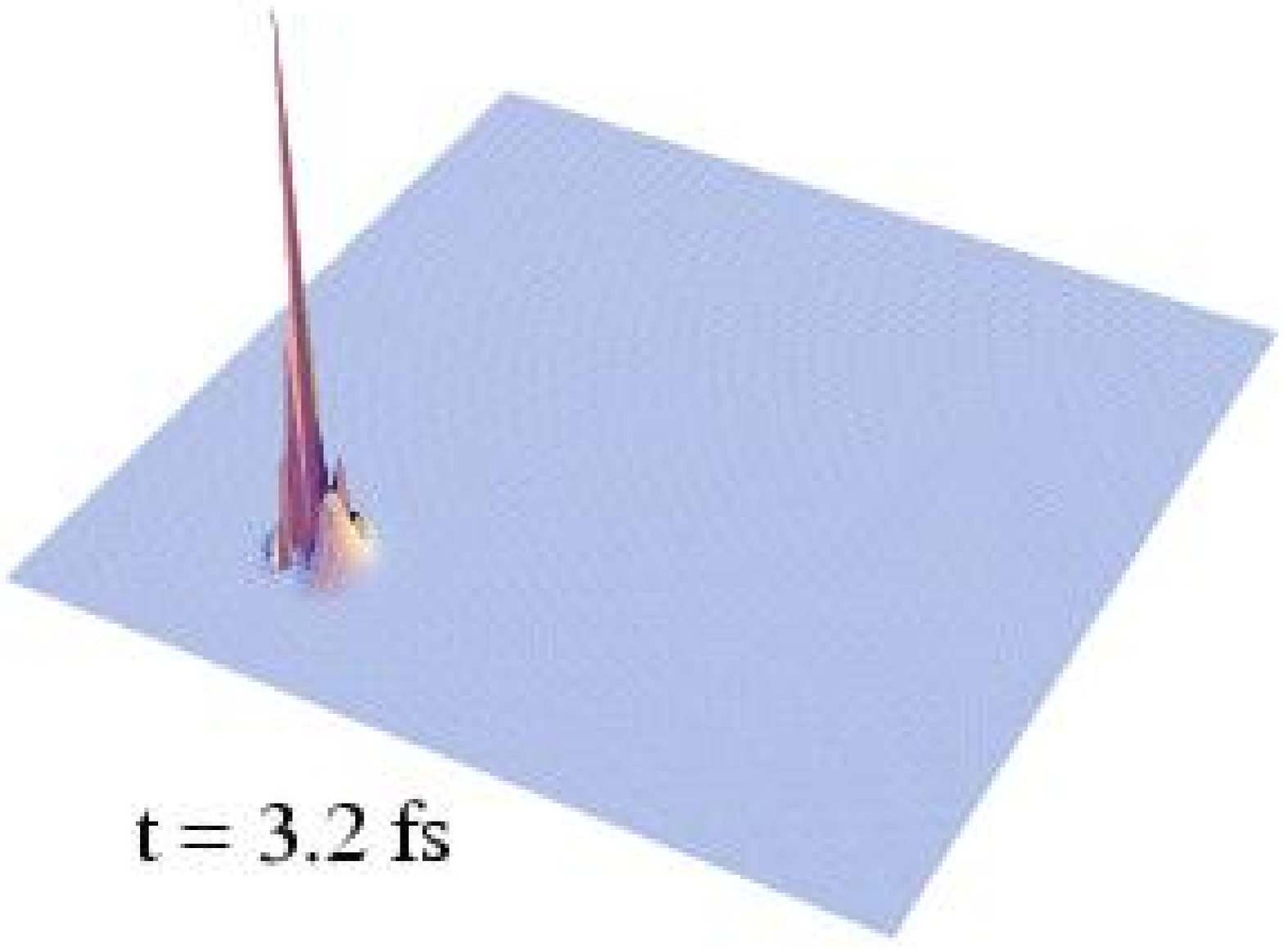}
    \includegraphics[clip,width=6cm,bbllx=0,bblly=0,bburx=590,bbury=480]{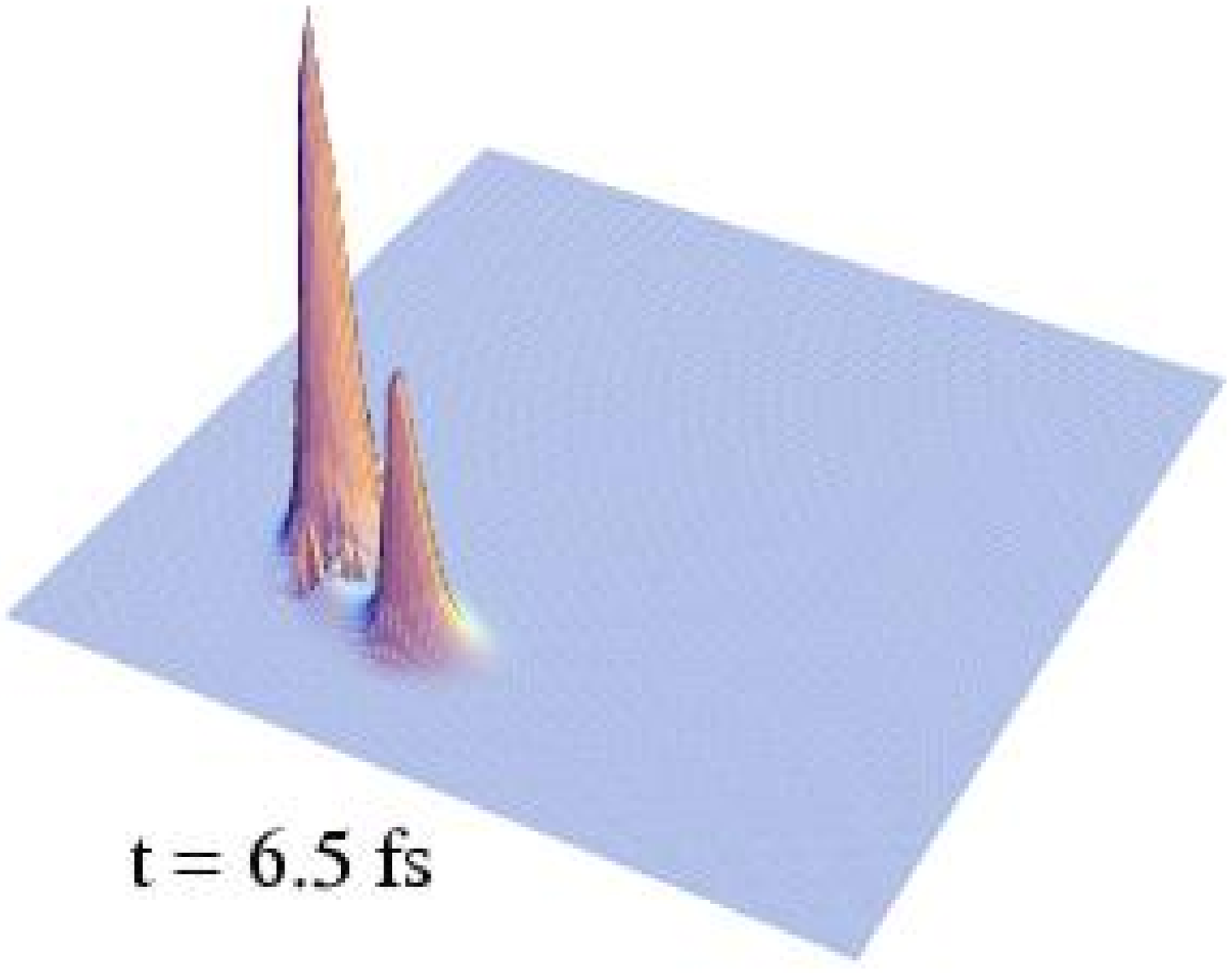}
    \includegraphics[clip,width=6cm,bbllx=0,bblly=0,bburx=590,bbury=480]{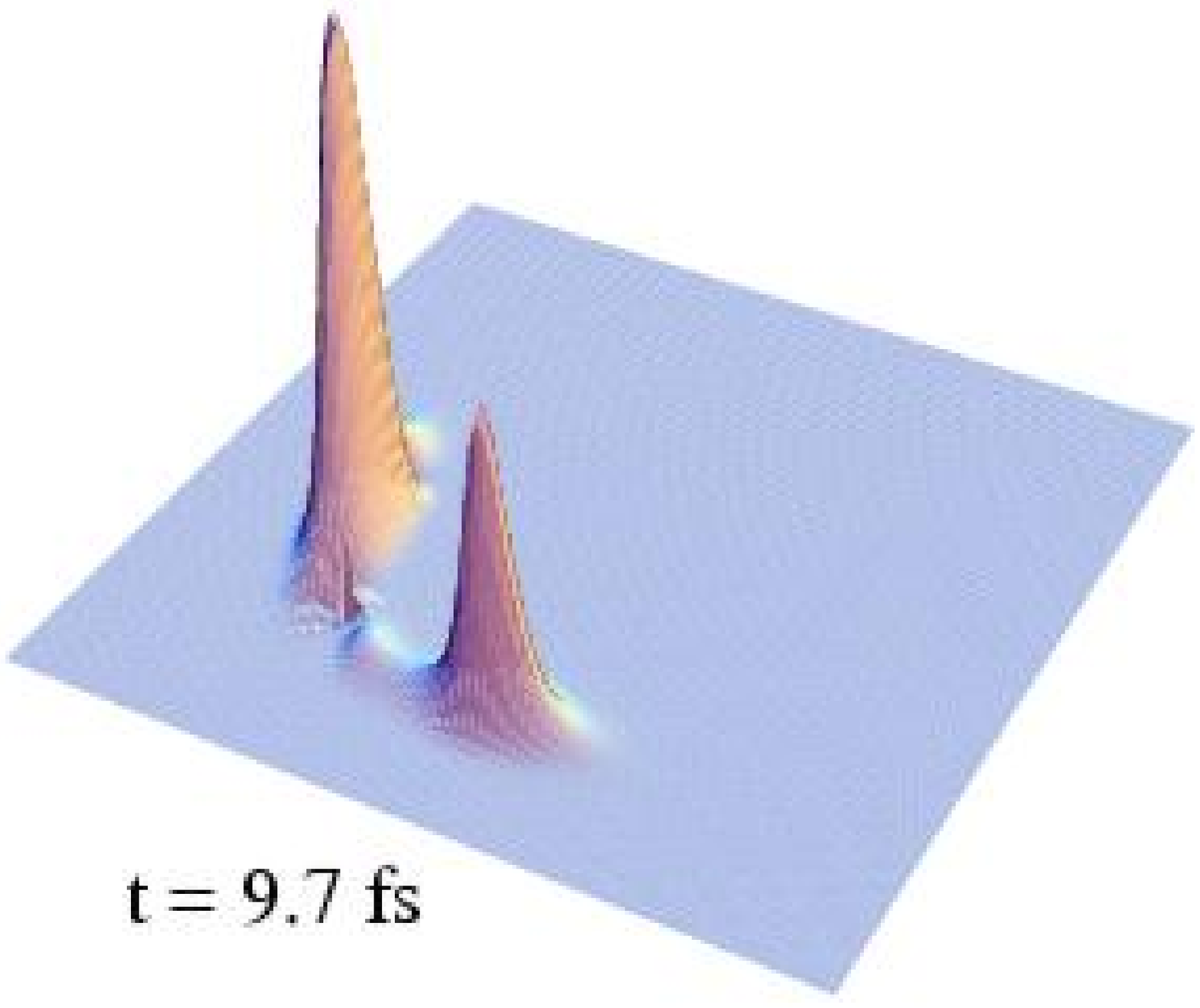}
    \includegraphics[clip,width=6cm,bbllx=0,bblly=0,bburx=590,bbury=480]{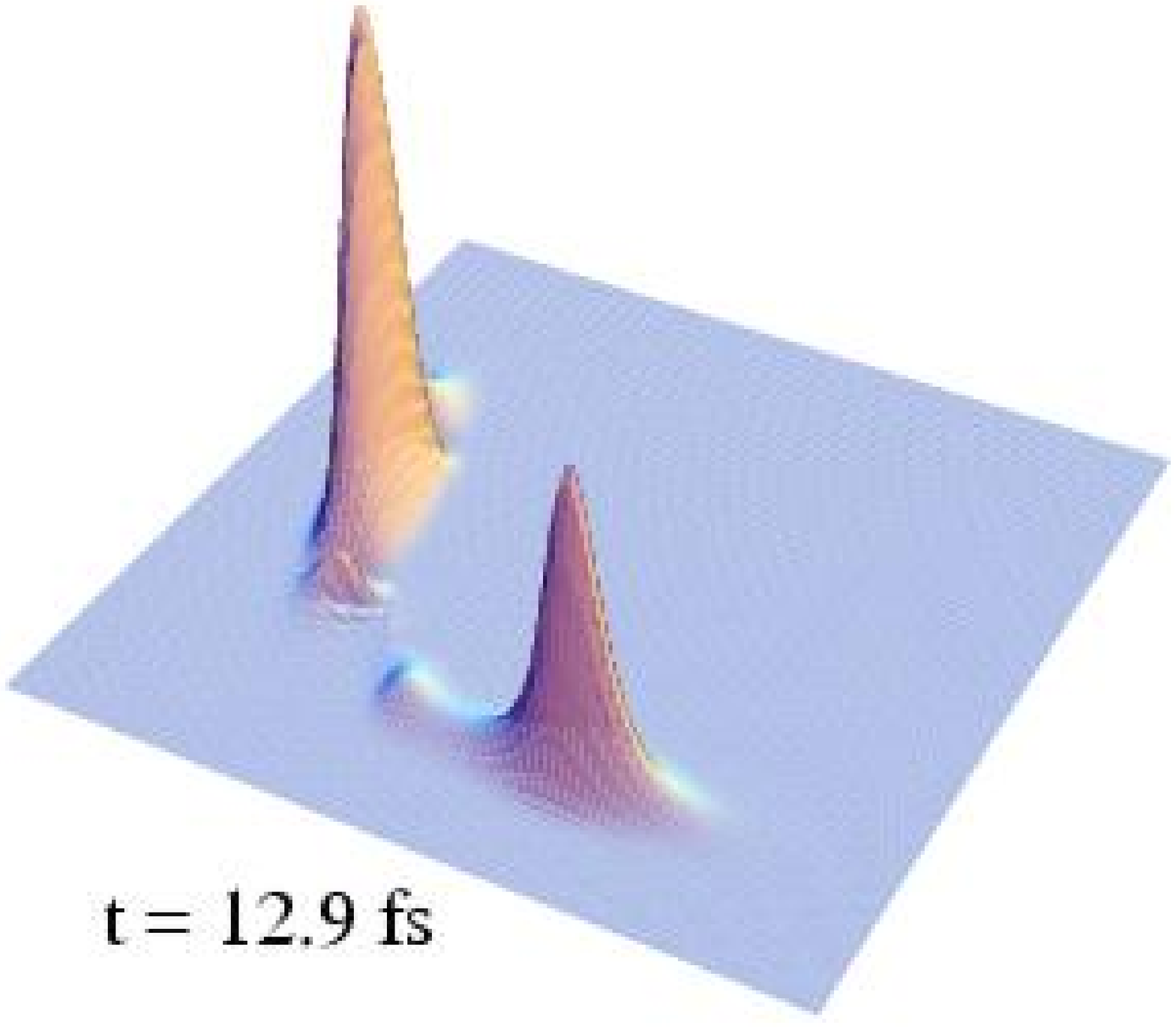}
    \includegraphics[clip,width=6cm,bbllx=0,bblly=0,bburx=590,bbury=480]{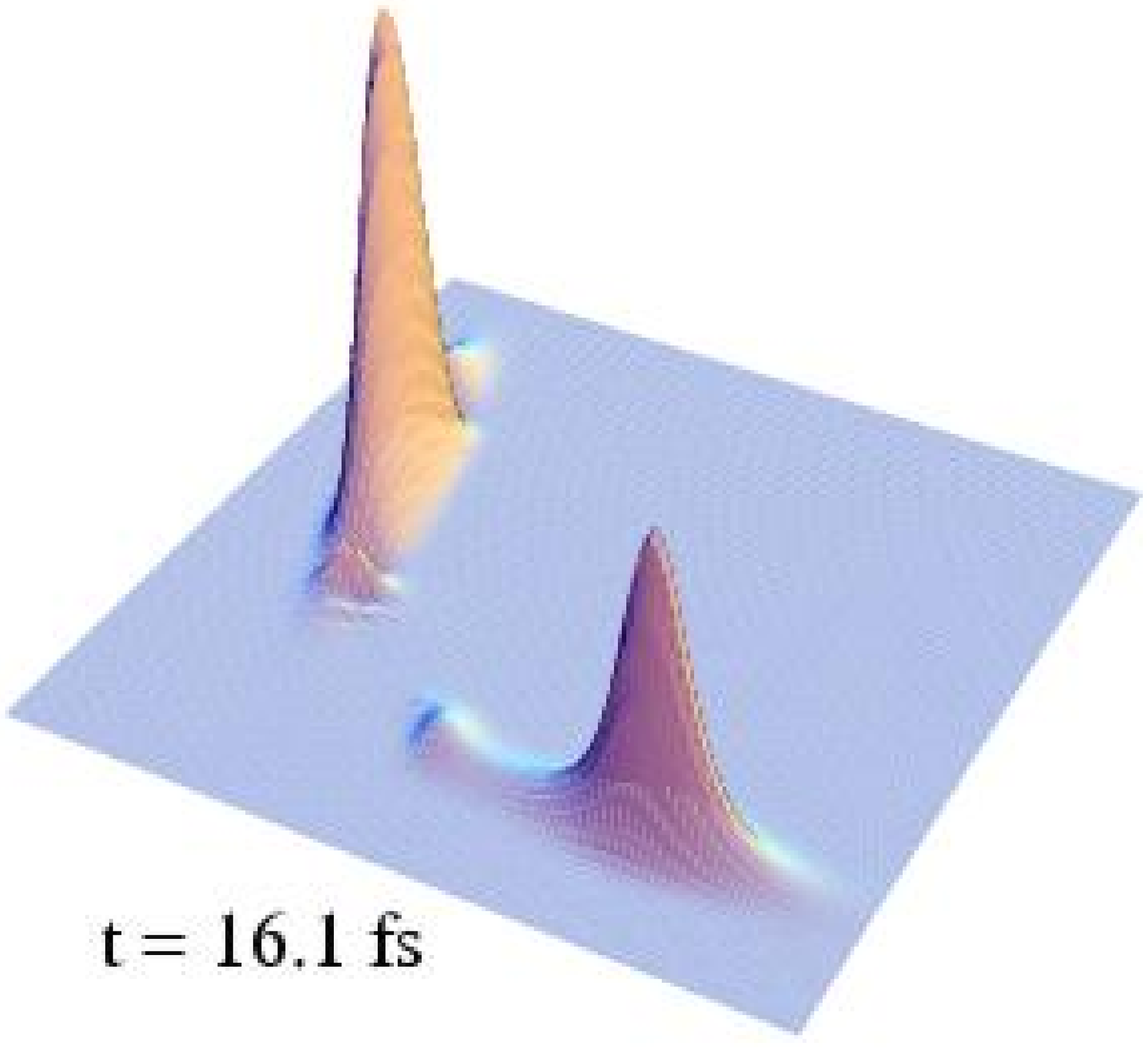}
    \caption{Snapshots of the broad-energy wave packet scattering from the double barrier potential
    with $\alpha=52^{\circ}$.}
    \label{fig.prop-45}
\end{figure}

The incident ballistic electrons are modelled using a Gaussian
wave packet
\begin{equation}
\psi(x,y,0)= exp((-(x -
L)^{2}/2\sigma_{x}^{2}-y^{2}/2\sigma_{y}^{2})+ i(x
\overline{p}_{x}+y \overline{p}_{y})/\hbar)\text{ ,}
\end{equation}
where $\sigma_{x}=\hbar/\sqrt{2}\Delta p_{x}$ and
$\sigma_{y}=\hbar/\sqrt{2}\Delta p_{y}$ are the wave packet's
standard deviation in the $x$ and $y$ dimensions, $\Delta p_{x}$
and $\Delta p_{y}$ are momentum uncertainty components,
$\overline{p}_{x}=\sqrt{2m^{\ast}\overline{\mathcal{E}}}$ and
$\overline{p}_{y}=0$ are the mean momentum components of the wave
packet with mean energy $\overline{\mathcal{E}}$. In the
simulations we consider electron wave packets with effective mass
$m^{\ast}=0.06m_{e}$ and $\overline{\mathcal{E}}$ in the range of
$[0.1, 3.5]V_{0}$. The narrow-energy wave packets (used in the
direct method) have a spacial distribution given by
$\sigma_{x}=\sigma_{y}=7.5$ nm corresponding to $\Delta E$ in the
range of $[0.02, 0.1]V_{0}$, and $L = 45.0$ nm. The broad-energy
wave packets (used in the momentum space method) are characterized
by $\sigma_{x}=0.75$ nm and $\sigma_{y}=7.5$ nm, mean energy
$\overline{\mathcal{E}}=1V_{0}$, and $L = 26.5$ nm. Wave packets
are propagated for a time $t=mD/p_{x}$ where $D\approx3L$ is the
propagation distance. The choice of parameters $L$ and $D$ ensures
that both the initial and final wave functions are sufficiently
away from the potential barriers (i.e.
$(\opamp{\psi}{\hat{V}}{\psi}/\opamp{\psi}{\hat{H}}{\psi} <
10^{-4})$). Fig. \ref{fig.prop-00} and \ref{fig.prop-45}
illustrate the position space evolution of the broad-energy wave
packet for the double barrier potentials with $\alpha=90^{\circ}$
and $52^{\circ}$.

\begin{figure}[h]
    \centering
    \includegraphics[width=12cm,bbllx=90,bblly=0,bburx=570,bbury=320]{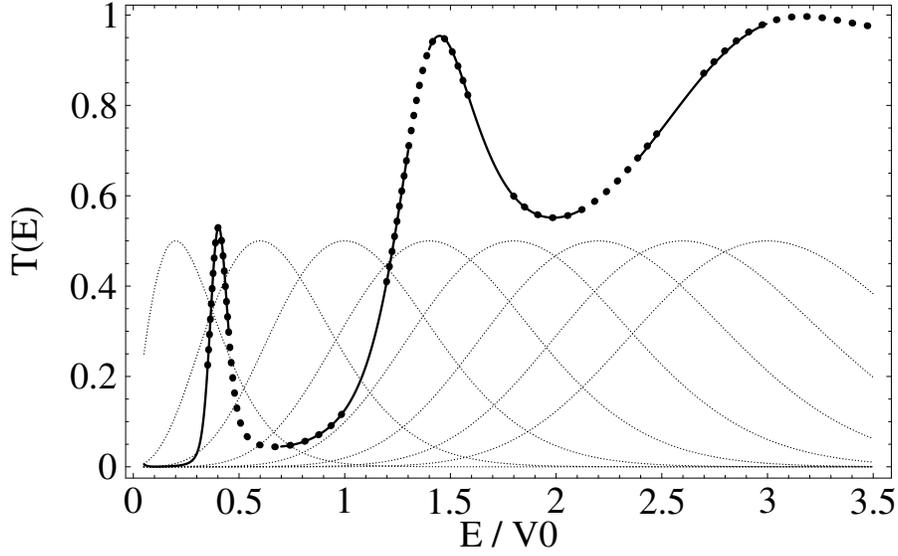}\\
    \caption{Continuous transmission spectrum for the double barrier potential
    with $\alpha=52^{\circ}$. Eight broad-energy wave packets with
    mean energies $\overline{\mathcal{E}}/V_{0}=0.2, 0.6, 1.0, 1.4, 1.8, 2.2,
    2.6,$ and  $3.0$ have been used to obtain the corresponding segments
    of the final curve. The agreement between these overlapping segments
    demonstrates that the transmission curve is independent of the
    choice of $\overline{\mathcal{E}}$.}
    \label{fig.multi-trans}
\end{figure}

In order to examine the accuracy of the proposed momentum space
method, we show firstly that the transmission curves are
independent of the choice of $\overline{\mathcal{E}}$ for the
broad-energy wave packets. Fig. \ref{fig.multi-trans} presents a
transmission curve for the double barrier potential with
$\alpha=52^{\circ}$, constructed by overlapping multiple segments
from transmission curves obtained using initial wave packet with
$\overline{\mathcal{E}}/V_{0}=0.2, 0.6, 1.0, 1.4, 1.8, 2.2, 2.6,$
and $3.0$ respectively. An excellent match between these segments
producing a uniform transmission spectrum is demonstrated.

\begin{figure}[h]
    \centering
    \includegraphics[width=12cm,bbllx=90,bblly=0,bburx=570,bbury=340]{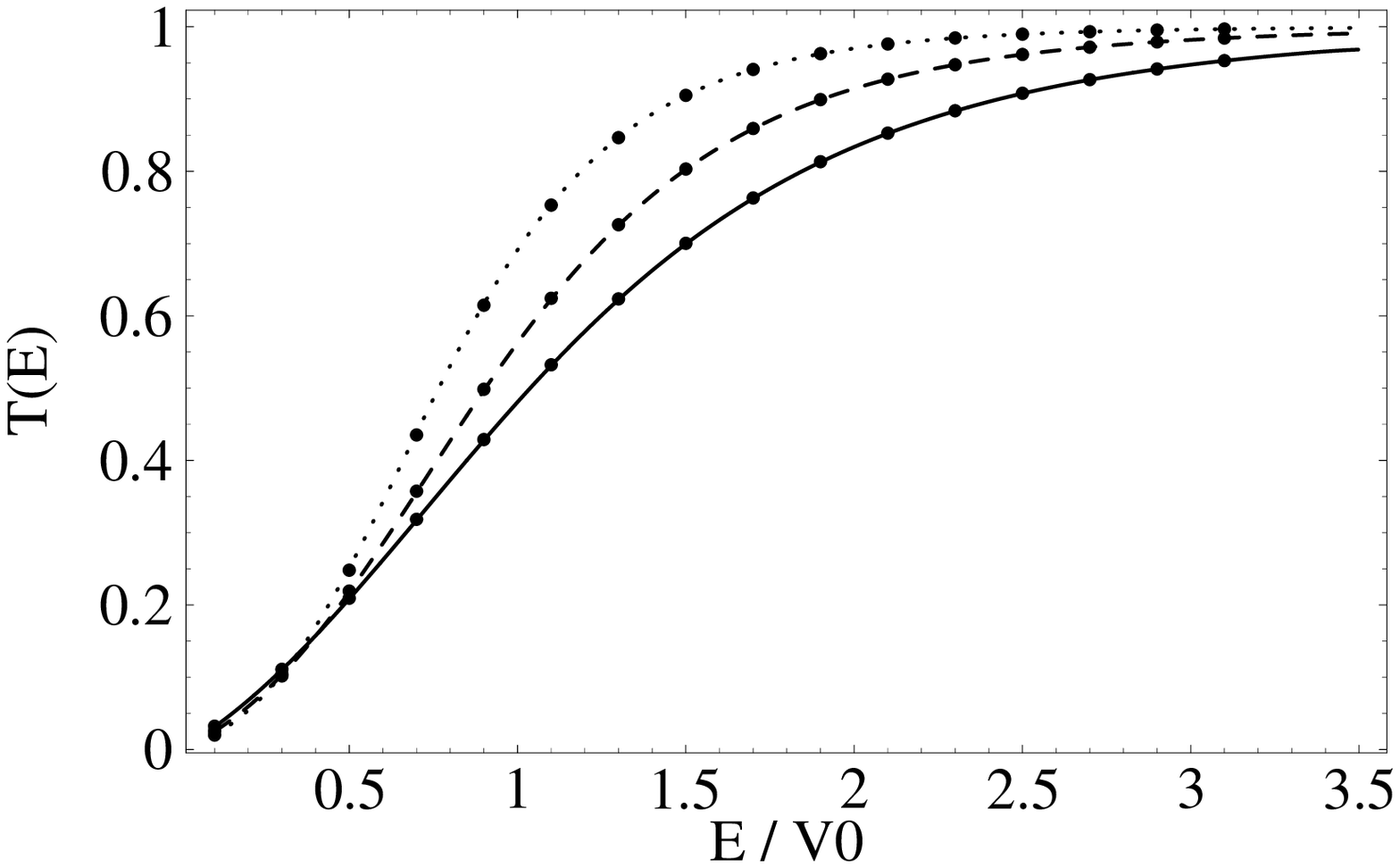}
    \caption{Continuous transmission spectra for the single barrier potential
    with orientation angles $\alpha=$ $52^{\circ}$ (solid), $62^{\circ}$ (dashed)
    and $90^{\circ}$ (dotted). Points on each curve represent the
    corresponding transmission coefficients calculated using the direct method.}
    \label{fig.trans-sb}
\end{figure}

\begin{figure}[h]
    \centering
    \includegraphics[width=12cm,bbllx=90,bblly=0,bburx=570,bbury=340]{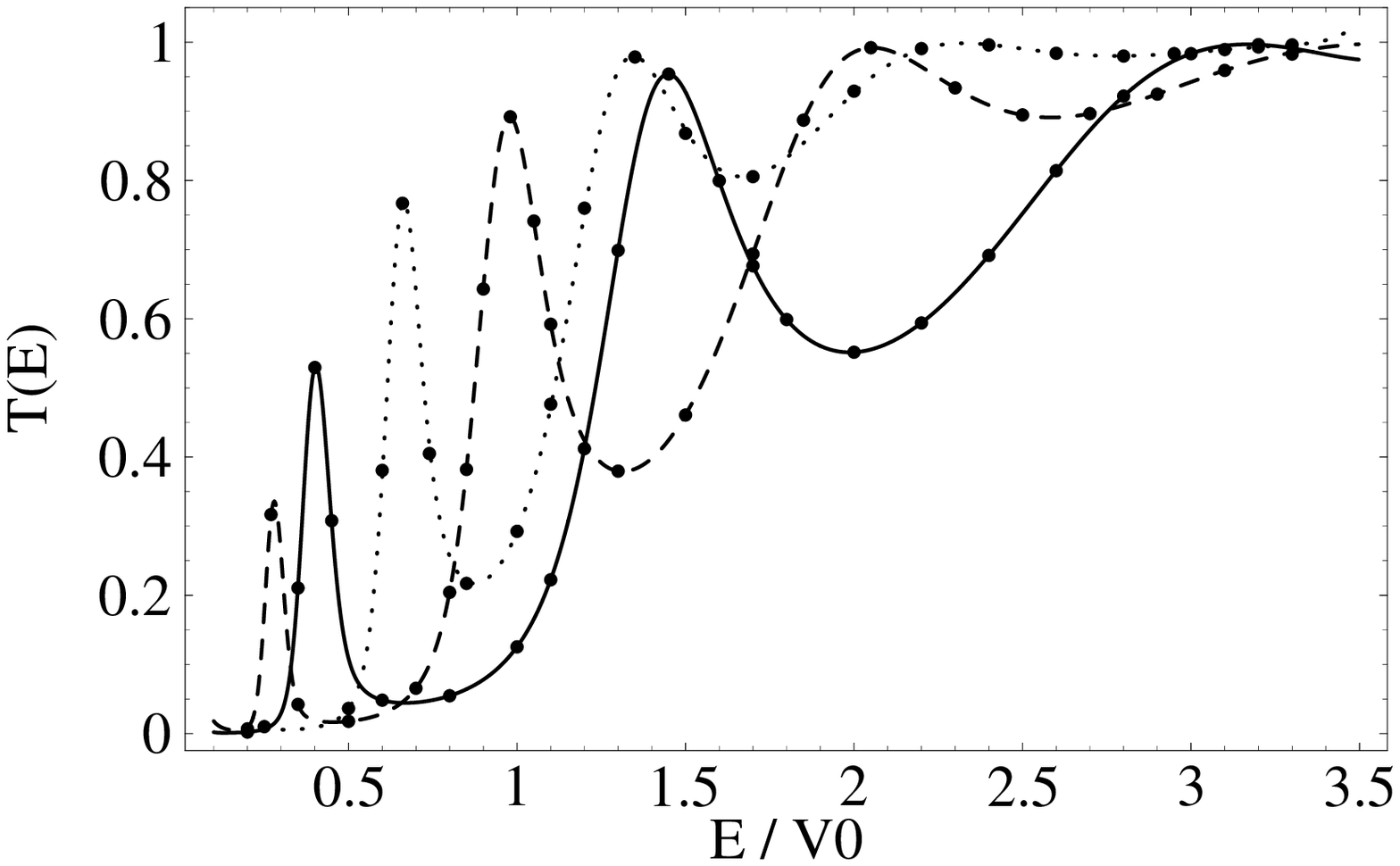}
    \caption{Continuous transmission spectra for the double barrier potential
    with orientation angles $\alpha=$ $52^{\circ}$ (solid), $62^{\circ}$ (dashed)
    and $90^{\circ}$ (dotted). Points on each curve represent the
    corresponding transmission coefficients calculated using the direct method.}
    \label{fig.trans-db}
\end{figure}

Secondly, we compare the continuous transmission curves obtained
using the momentum space method, against the discrete transmission
coefficients obtained using the direct method. Fig.
\ref{fig.trans-sb} and \ref{fig.trans-db} depict the results for
the single and double barrier potentials respectively, with three
orientation angles $\alpha=$ $52^{\circ}$, $62^{\circ}$ and
$90^{\circ}$. A close match between each transmission curve and
its corresponding transmission coefficients (points on each curve)
demonstrates an excellent agreement between the two methods.

\begin{figure}[h]
    \centering
    \includegraphics[width=11.5cm,bbllx=90,bblly=0,bburx=570,bbury=320]{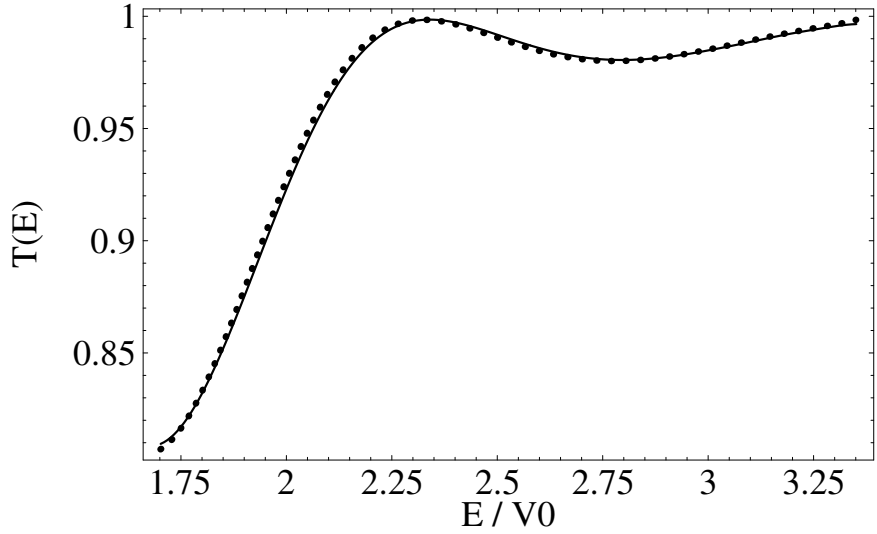}
    \caption{Transmission curves for the double barrier potential
    with $\alpha=52^{\circ}$, obtained using the accuracy factor
    $\beta=1.0$ (solid) and $\beta=1.2$ (dotted).
    The agreement between the two curves demonstrates that
    good convergence has already been achieved at $\beta=1.0$.}
    \label{fig.trans-comp-1}
\end{figure}

\begin{figure}[h]
    \centering
    \includegraphics[width=11.5cm,bbllx=90,bblly=0,bburx=570,bbury=320]{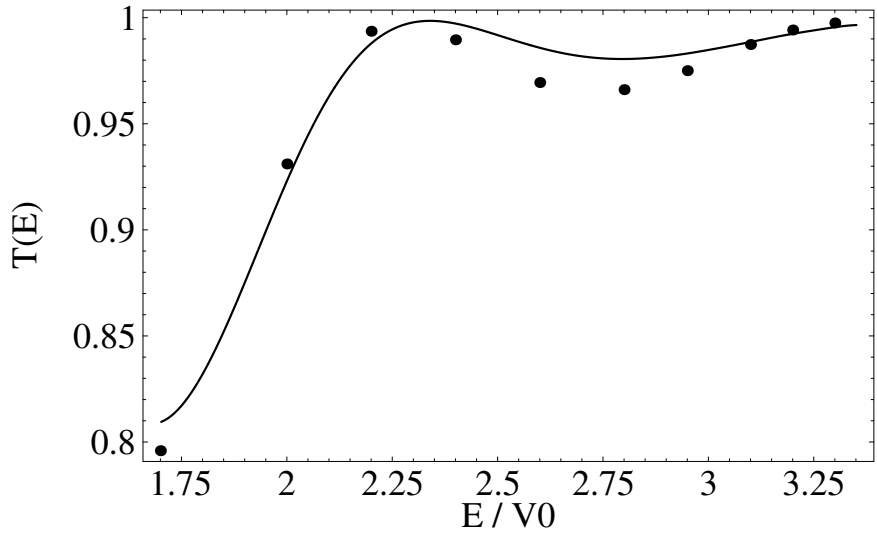}
    \caption{Transmission curve for the double barrier potential
    with $\alpha=52^{\circ}$ vs. direct method transmission points.
    Both results have been obtained using
    the accuracy factor $\beta=1.0$. Differences between
    the transmission curve and its corresponding transmission points suggest that
    the direct method does not achieve complete convergence at $\beta=1.0$.}
    \label{fig.trans-comp-2}
\end{figure}

\begin{figure}[h]
    \centering
    \includegraphics[width=11.5cm,bbllx=90,bblly=0,bburx=570,bbury=320]{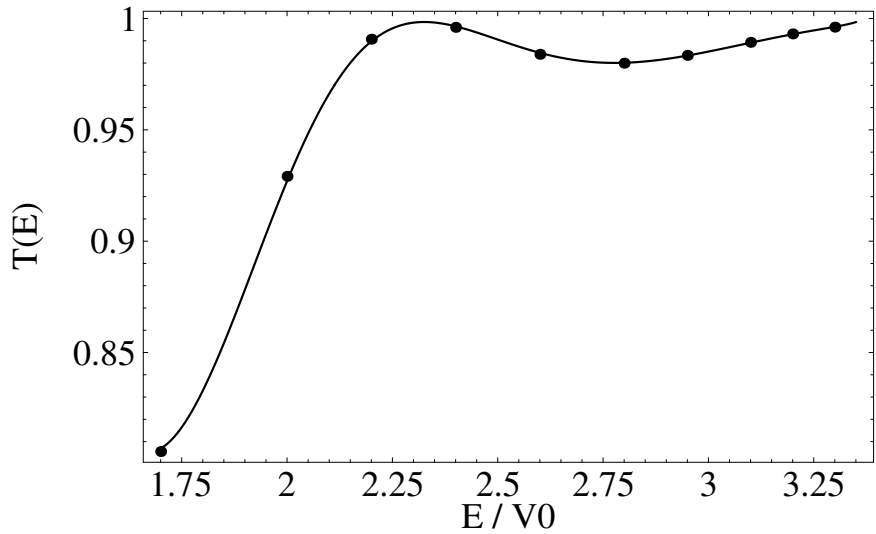}
    \caption{Transmission curve for the double barrier potential
    with $\alpha=52^{\circ}$ vs. direct method transmission points.
    By increasing the accuracy factor $\beta$ of the direct method from $1.0$ to $2.0$,
    we arrive at an agreement between the transmission curve and
    its corresponding transmission points. This suggests that
    the direct method requires $\beta>1.0$ for complete
    convergence.}
    \label{fig.trans-comp-3}
\end{figure}

In order to examine the computational efficiency of the momentum
space method, we first note that the accuracy and convergence of
the Chebyshev-Fourier propagation scheme primarily depends on the
choice of $\mathcal{E}_{max}$ in Eq. (\ref{eqn.Chebyshev}), which
determines the dimensions of the numerical grid due to Nyquist's
sampling theorem. This accuracy improves asymptotically with
higher values of $\mathcal{E}_{max}$ while making computation more
costly. In the simulations $\mathcal{E}_{max}$ is estimated based
on the maximum potential energy of the system $V_{0}$, the mean
momentum of the initial wave packet, and its maximum momentum
range. We then define an accuracy factor $\beta$ to be the ratio
between the actual value of $\mathcal{E}_{max}$ used in the
simulation and the initial estimation. All results from the
momentum space method were obtained using $\beta=1.0$, which was
found to provide good accuracy while maintaining a reasonable
computational speed. The direct method however, at times, required
a higher $\beta$ value for convergence. In the case of the double
barrier with $\alpha=90^{\circ}$ for instance, increasing $\beta$
from $1.0$ to $1.2$, produces no discernable improvements in the
resulting transmission curve using the momentum space method (Fig.
\ref{fig.trans-comp-1}). Using the direct method however, setting
$\beta=1.0$ leads to significant discrepancies between the
transmission points and the transmission curve (Fig.
\ref{fig.trans-comp-2}). Conversely, setting $\beta=2.0$ results
in an excellent agreement (Fig. \ref{fig.trans-comp-3}), but
severely increases the cpu time requirement.

The efficiency of the momentum space scheme becomes particularly
apparent when considering the existence of narrow transmission
peaks in the spectrum (such as the one at
$\overline{\mathcal{E}}=0.4V_{0}$ for the double barrier potential
with $\alpha=52^{\circ}$). In the above simulations, obtaining a
continuous transmission curve using the momentum space method was
$5$ times faster than the application of the direct method, when
requiring only $20$ points to represent the same curve. In the
absence of prior information about the topography of the
transmission spectrum however, using the direct method
necessitates probing the spectrum with a much higher resolution,
in order to detect possible narrow transmission peaks. As such,
applying the direct method can become prohibitively inefficient,
rendering the use of momentum space scheme vastly advantageous.

\section{Conclusion}

We presented an efficient scheme for computing the continuous
transmission spectrum of charge carriers scattered by an arbitrary
potential geometry in two dimensions. We applied this scheme to
single and double barrier potentials at various angles of
incidence, and demonstrated excellent precision as well as
substantial computational saving as compared to a standard method
commonly used for calculating transmission coefficients.

%===========================================
% BIBLIOGRAPHY
%===========================================
\bibliography{ref}

%===========================================
% FIGURES
%===========================================
\newpage
\section{Figures}

\end{document}